# Profiles of static liquid-gas interfaces in axisymmetrical containers under different accelerations

Shangtong Chen, Yong Gao, Wen Li, Fenglin Ding, Jintao Liu, Yong Li*

Beijing Institute of Control Engineering, China Academy of Space Technology, Beijing 100094, China

*Corresponding author: li-y95@tsinghua.org.cn

**Abstract**

Second-order perturbation solutions of profiles of bubbles suspended in liquid and liquid-gas interfaces when liquid all sinks in the bottom under different accelerations are derived. Six procedures are developed based on these solutions, and they are divided into two types. One takes coordinates of endpoints of profiles as inputs, and the other takes liquid volume or gas volume as inputs. Numerical simulation are performed with the Volume of Fluid method and numerical results are in good agreement with predictions of these procedures. Besides, the bigger the acceleration, the more flatter the bubble will be until all liquid sinks to the bottom. Effects of accelerations on bubbles' shape must be considered, otherwise it will cause a volume error of about 10%. When liquid all sinks to the bottom, predictions of liquid volume with the same liquid meniscus height as inputs differs a lot under different accelerations. The most significant change of liquid volume is when *Bond* << 1. Effects of accelerations and liquid contact angle on liquid-gas interfaces must be considered during evaluating liquid residue, and these findings will be great helpful for liquid residue measurement and fine management in space.

**Keywords:** Liquid-gas interface, axisymmetrical container, different accelerations, propellant residue

# 1. Introduction

The residual microgravity in space and the accelerations caused by spacecraft maneuvering have significant effects on the morphology of liquid-gas interfaces in tanks. It's very important to accurately predict the morphology of static liquid-gas interfaces in tanks under different accelerations for fluid management and liquid residue measurement in space.

Liquid in tanks is mainly affected by the surface tension, satellite maneuvering accelerations and residual microgravity. Surface tension driven flows have been widely explored. Weislogel et al (1998) derived dynamic equations of surface tension driven flows along interior corners, and the same method is adopted to derive the dynamic equations of flows along interior corners formed by planar walls of varying wettability (Weislogel et al., 2005), rounded interior corners (Chen et al., 2006), interior corners of rounded wall (Li et al., 2015), curved interior corners (Wu et al., 2018) and interior corners in the plate tanks (Zhuang et al., 2012). Zhou et al. (2020) comprehensively studied surface tension driven flows along corners with arbitrary cross-sections and presented the universal mathematical models of flows. Tian et al. (2019) derived dynamic equations of liquid climbing in a narrow and tilting corner by seeking the minimum of the Rayleighian. Surface tension driven flows in tubes are also widely studied , including cylindrical tubes (Stange et al., 2003), oval tubes (Chen et al., 2021), rectangular channels (Wang et al. 2021), tubes with corners (Zhao et al., 2022), injector tubes of monopropellant thrusters (Chen et al., 2013), eccentric annuli (Chen et al. 2023a). Flows in tubes with varying cross-sections (Lei et al., 2021; Figliuzzi et al., 2013) are explored and the methods to optimize the channels are proposed.

Static liquid-gas interfaces under microgravity and normal gravity are also deeply analyzed. Theoretical expressions of profiles of liquid drops on the walls of revolutions under microgravity are derived from the Young-Laplace equation (Carroll, 1976; Michielsen et al., 2011; Du et al., 2010, 2011; Chen et al., 2022, 2023b, 2023c). Michielsen et al. (2011) obtained the static position and free surfaces pf liquid drops in conical fibers by seeking the minimum free surface energy. Chen et al. (2022, 2023b, 2023c) uses the shooting method to predict the free surfaces when given liquid volume. Theoretical expressions of profiles of capillary bridges between different structures under microgravity are also derived from the Young-Laplace equation (Mason et al., 1965; Clark et al., 1968; Fortes, 1982; Honschoten et al., 2010; Wang et al., 2013; Timothy et al.; 2015; Reyssat, 2015).

Besides capillary surfaces in absence of gravity, much attention are also paid to free surfaces under different accelerations. Concus (1968) established the mathematical models of free surfaces in cylinders by using the perturbation method. The same method is used to derive the expressions of of sessile drops (Chesters, 1977; O'Brien et al., 1991; Shanahan, 1982; Rienstra, 1990; Yariv et al., 2022, 2023). The minimum energy method are also used to derive expressions of profiles of sessile drops (Shanah, 1984). Padday et al. (1997) explored the shape, stability and breakage of pendant liquid bridges by balancing the gravity and surface tension. Smith et al. (1984) studied axisymmetric drops and developed an asymptotic solution to the Laplace equation which is valid when the ratio of gravity to surface tension forces is small.

However, static free surfaces in axisymmetrical containers, such as propellant tanks, under different accelerations haven't been studied deeply. There may be a big bubble suspended in liquid, or all liquid sinks to the bottom under different accelerations. The free surface morphology, deformation rules and prediction methods have not been obtained. Effects of the geometry, acceleration and liquid contact angle on the morphology of liquid-gas interfaces are not clear. In this paper, the bubble morphology and the free surfaces when liquid all sinks in the bottom under different accelerations are deeply explored, and theoretical expressions of profiles of these free surfaces are established through the perturbation method. Influences of accelerations, geometries, liquid contact angle, liquid filling rate and other factors on the free surfaces are revealed. Based on these theoretical models, by combining with the shooting method procedures are developed to predict free surfaces under different conditions. Numerical simulation is carried out with the Volume of Fluid (VOF) method by considering different liquid contact angles, different geometries and other factors, and the numerical results are compared with predictions of the procedures.

## 2. Bubbles suspended in liquid

**2.1 Theoretical derivation**

When accelerations can be ignored, bubbles in tanks are spherical and locates in the middle region. When the acceleration plays a role but the *Bond* (*Bo*) number is small enough, the bubble's shape will deviate from the spherical shape, and the final stable position will deviate in the opposite direction of the acceleration. The more obvious the acceleration effect is, the greater the deviation will be until all the liquid sinks to the bottom of the container.

Figure 1 shows liquid distribution in the half cross-section of the model when the *Bo* number is small enough, in which the blue and white regions represent the liquid and gas respectively. The tank can be of any shape, and for convenience the spherical shape is adopted for analysis, whose radius is $r_t$. The cylindrical coordinate *rOz* is used in this study. The original point *O* locates in the highest point of bubbles' profiles, which will make the expression of static pressure on the liquid-gas interface easier. The *z*-axis coincides with the bubble's symmetry axis, and the *r*-axis is perpendicular to the *z*-axis and horizontally to the right. Point *A* ($r_1$, $z_1$) is the point with the largest abscissa on the profile, and Point *B* (0, $z_2$) is the point with the largest ordinate on the profile. $\varphi$ is the inclination angle between the profile of liquid-gas interfaces and the positive direction of *r*-axis. *s* is the length of the profile measured from Point *O*. Because the liquid-gas interface is axisymmetrical with respect to *z*-axis, the profile in the first quadrant is adopted in the theoretical derivation. Some basic assumptions are presented as follows:

1. The liquid is Newtonian, incompressible, and homogeneous.

2. The liquid density is much larger than that of the gas so that effects of the gas density can be ignored.

3. No stress acts on liquid-gas interfaces, so the interface is decided only by the acceleration and the surface tension.

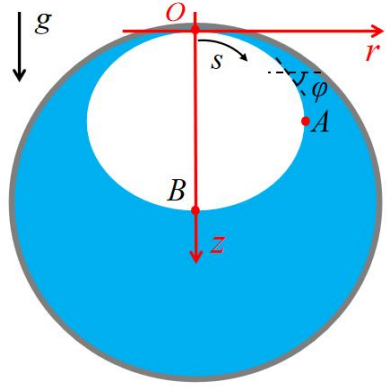

**Fig. 1** Hal cross-sectional view of the simplified model.

In most cases the distance between the highest point on the bubble's profile (Point $O$) and the highest point of the container is very small as shown in Fig. 1, and the static pressure caused by this distance can be ignored when the bubble's size is much larger than the distance. The hydrostatic pressure on the profile of the liquid-gas interface is given by

$$p = \rho g z + p_O \tag{2.1}$$

where $\rho$ is liquid density, $g$ is the acceleration and $p_O$ is the pressure jump across the curved free surface at $z = 0$. On static liquid-gas interfaces the hydrostatic pressure is balanced by the capillary pressure, which leads to the following equilibrium

$$\sigma(K_1 + K_2) = \rho g z + p_O \tag{2.2}$$

where $\sigma$ is liquid surface tension, $K_1$ and $K_2$ are the principal curvature of the profile. Combined with their expressions in $rOz$ coordinates Eq. (2.2) can be expressed as follows

$$\sigma\left(\frac{d^2z/dr^2}{[1+(dz/dr)^2]^{1.5}} + \frac{dz/dr}{r[1+(dz/dr)^2]^{0.5}}\right) - \rho g z = p_O \tag{2.3}$$

With the following relationship

$$\frac{dr}{ds} = \cos\varphi \tag{2.4}$$

$$\frac{dz}{ds} = \sin\varphi \tag{2.5}$$

Eq. (2.3) are transformed into

$$\sigma\left(\frac{d\varphi}{ds} + \frac{\sin\varphi}{r}\right) - \rho g z = p_O \quad (2.6)$$

which becomes a first-order equation. Combined with Eqs. (2.4)-(2.6), $r$ and $z$ can be expressed as functions of the inclination angle $\varphi$

$$\frac{dr}{d\varphi} = \frac{\sigma r \cos\varphi}{-\sigma \sin\varphi + \rho g r z + p_O r} \quad (2.7)$$

$$\frac{dz}{d\varphi} = \frac{\sigma r \sin\varphi}{-\sigma \sin\varphi + \rho g r z + p_O r} \quad (2.8)$$

The boundary conditions are listed as follows

$$\varphi = 0, \quad r = 0, z = 0, s = 0 \quad (2.9)$$

$$\varphi = \frac{\pi}{2}, \quad r = r_1, z = z_1 \quad (2.10)$$

If the *Bo* number is 0 the bubble is spherical and this is the basis for perturbation solutions. The small *Bo* number (*Bo* << 1) is needed to derive the perturbation solution. Eqs. (2.7)&(2.8) can be nondimensionalized by using capillary length, $l = \sqrt{\sigma/\rho g}$, and $r_1$, which leads to

$$\frac{dX}{d\varphi} = \frac{X\cos\varphi}{-\sin\varphi + \beta XY + XP_*} \quad (2.11)$$

$$\frac{dY}{d\varphi} = \frac{X\sin\varphi}{-\sin\varphi + \beta XY + XP_*} \quad (2.12)$$

where $X = r/r_1, Y = z/r_1, P_* = p_o r_1/l, \beta = \frac{\rho g r_1^2}{\sigma}$.

The scaled boundary conditions are

$$\varphi = 0, \quad X = 0, Y = 0, S = 0 \quad (2.13)$$

$$\varphi = \frac{\pi}{2}, \quad X = 1, Y = z_1/r_1 \quad (2.14)$$

When $\beta$ << 1, according to the perturbation method the solutions of Eqs. (2.11)&(2.12) are expressed as

$$X = X_0 + \beta X_1 + \beta^2 X_2 \quad (2.15)$$

$$Y = Y_0 + \beta Y_1 + \beta^2 Y_2 \tag{2.16}$$

$$P_* = P_{*0} + \beta P_{*1} + \beta^2 P_{*2} \tag{2.17}$$

Inserting the zero-order solutions of Eqs. (2.15)-(2.17) into Eqs. (2.11)&(2.12), we obtain that

$$\beta X_0 X_0' Y_0 + X_0 X_0' P_{*0} - X_0' \sin\varphi = X_0 \cos\varphi \tag{2.18}$$

$$\beta X_0 Y_0 Y_0' + X_0 Y_0' P_{*0} - Y_0' \sin\varphi = X_0 \sin\varphi \tag{2.19}$$

where $X_0'$ and $Y_0'$ represents $dX_0/d\varphi$ and $dY_0/d\varphi$ respectively. Because $\beta \ll 1$ and the terms $O(\beta)$ are ignored. It can be obtained that

$$X_0 = \sin\varphi \tag{2.20}$$

$$Y_0 = 1 - \cos\varphi \tag{2.21}$$

$$P_{*0} = 2 \tag{2.22}$$

Similarly, inserting the first-order solutions of Eqs. (2.15)-(2.17) into Eqs. (2.11)&(2.12), ignoring the terms $O(\beta^2)$, and simplifying the results by combining Eqs. (2.20)-(2.22) lead to

$$X_0 Y_0 X_0' + X_0 P_{*1} X_0' + X_1 P_{*0} X_0' + X_0 P_{*0} X_1' - X_1' \sin\varphi = X_1 \cos\varphi \tag{2.23}$$

$$X_0 Y_0 Y_0' + X_0 P_{*1} Y_0' + X_1 P_{*0} Y_0' + X_0 P_{*0} Y_1' - Y_1' \sin\varphi = X_1 \sin\varphi \tag{2.24}$$

where $X_1'$ and $Y_1'$ represents $dX_1/d\varphi$ and $dY_1/d\varphi$ respectively. According to Eqs. (2.13)&(2.14) the boundary conditions are written as

$$\varphi = 0, \ X_1 = 0, Y_1 = 0 \tag{2.25}$$

$$\varphi = \frac{\pi}{2}, X_1 = 0 \tag{2.26}$$

By using the same method adopted during deriving the zero-order solutions and combing Eqs. (2.20)-(2.22), it can be obtained that

$$X_1 = \frac{1}{3}\cos^2\varphi \tan\frac{\varphi}{2} \tag{2.27}$$

$$Y_1 = \frac{1}{3}\cos\varphi + \frac{1}{2}\sin^2\varphi + \frac{1}{6}\cos^2\varphi + \frac{1}{3}\ln\left(2\cos^2\frac{\varphi}{2}\right) - \frac{1}{2} - \frac{1}{3}\ln 2 \qquad (2.28)$$

$$P_{*1} = -\frac{1}{3} \qquad (2.29)$$

Likewise, inserting the second-order solutions of Eqs. (2.15)-(2.17) into Eqs. (2.11)&(2.12) and simplifying the results lead to

$$X_2'\sin\varphi + X_2\cos\varphi = -\left(X_0 Y_1 X_0' + X_1 Y_0 X_0' + X_0 P_{*2} X_0' + X_1 P_{*1} X_0' + X_0 Y_0 X_1' + X_0 P_{*1} X_1' + X_1 P_{*0} X_1'\right)$$
(2.30)

$$Y_2'\sin\varphi + X_2\sin\varphi = -\left(X_0 Y_1 Y_0' + X_1 Y_0 Y_0' + X_0 P_{*2} Y_0' + X_1 P_{*1} Y_0' + X_0 Y_0 Y_1' + X_0 P_{*1} Y_1' + X_1 P_{*0} Y_1'\right)$$
(2.31)

and the solutions are obtained as follows

$$X_2 \sin\varphi = \frac{1}{12}\cos\varphi + \frac{1}{36}\cos 3\varphi - \frac{1}{8}\sin^4\varphi + \frac{1}{24}\cos^4\varphi - \left(\frac{1}{8} + \frac{1}{12}\ln 2\right)\cos 2\varphi - \frac{P_{*2}}{2\sin^2\varphi}$$
$$+ \frac{1}{12}\left\{\ln\left[2\cos^2\frac{\varphi}{2}\right]*(\cos 2\varphi - 1) + 2\cos\varphi - \cos^2\varphi\right\} - \frac{2}{3}\left[\sin^2\frac{\varphi}{2}\sin\varphi\cos^2\varphi\tan\frac{\varphi}{2} + \frac{1}{6}\cos^3\varphi - \frac{1}{8}\cos^4\varphi\right]$$
$$- P_{*0}\frac{X_1^2}{2} - P_{*1}X_0 X_1 - \frac{1}{12} + \frac{1}{12}\ln 2$$
(2.32)

$$Y_2 = -\frac{1}{6}\ln 2 + \cos\varphi\left(\frac{1}{6}\ln 2 + \frac{23}{18}\right) - \frac{5}{9}\cos^2\varphi + \frac{2}{9}\cos^3\varphi$$
$$- \frac{3\cos\varphi}{\sin^2\varphi}\left(-\frac{1}{27}\cos^4\varphi + \frac{1}{9}\cos^3\varphi - \frac{8}{27}\cos^2\varphi - \frac{2}{9}\cos\varphi + \frac{4}{9}\right) \qquad (2.33)$$
$$+ \ln\cos\frac{\varphi}{2}\left(-\frac{1}{3}\cos\varphi - \frac{2}{9}\right) - Y_0 Y_1 + \frac{Y_1}{3}$$

$$P_2 = \frac{1}{3}\ln 2 - \frac{1}{6} \qquad (2.34)$$

Second-order perturbation solutions of profiles of bubbles with small *Bo* numbers are now established.

**2.2 Numerical simulation**

Numerical simulation is performed in Fluent with the VOF method. To reduce the calculation time one half of the container is adopted. Fig. 2 shows a 3D mesh model. The grid density is increased near the container's wall by considering boundary effects. In the simulation, the first phase is the air and the second phase is a kind of silicone oil named by its kinematic viscosity (SF 10). Properties of fluids are shown in Table 1. Numerical settings in Fluent are shown in Table 2.

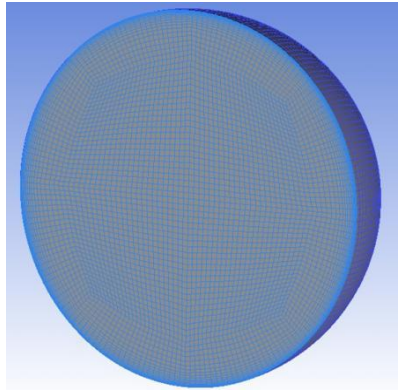

**Fig. 2** 3D mesh model of a hemisphere.

**Table 1** Fluid properties (25°C).

| Fluid | $\mu$ (kg/(m*s)) | $\rho$ (kg/m$^3$) | $\sigma$ (N/m) | $v$ ($10^{-6}$ m$^2$/s) |
|---|---|---|---|---|
| Air | 1.789e-5 | 1.225 | / | 1.460e-5 |
| SF 10 | 0.00935 | 935 | 0.0201 | 10 |

**Table 2** Numerical settings in Fluent.

| Name | Settings |
|---|---|
| Spatial discretization of the pressure equation | Body Force Weighted |
| Spatial discretization of the gradient equation | Least Square Cell |
| Spatial discretization of the momentum equation | Second-order Upwind Scheme |
| Spatial discretization of the volume fraction equation | Geo-Reconstruct |
| Pressure-velocity coupling equation | SIMPLEC |

In the beginning the bubble is spherical and in the middle region of containers, as shown in the left part of Fig. 3, the blue region represents the gas. Bubble volume is expressed as $V_b$. The middle and right part of Fig. 3 shows the static shapes of bubbles under 0.001 m/s$^2$ and 0.002 m/s$^2$ respectively. Under the influence of accelerations, bubble move to the upper part of the container and become flatter in shape. The bigger the acceleration, the more obvious the flattening of bubbles will be until all the liquid sinks to the bottom. Coordinates of Points *A* and *B* can are measured from numerical results and presented in Table 3. Point *O* and Point *B* locates in the highest and lowest position of the bubble respectively according to the simplified model shown in Fig. 1. Point *A* locates in the position whose abscissa is the largest on the profile.

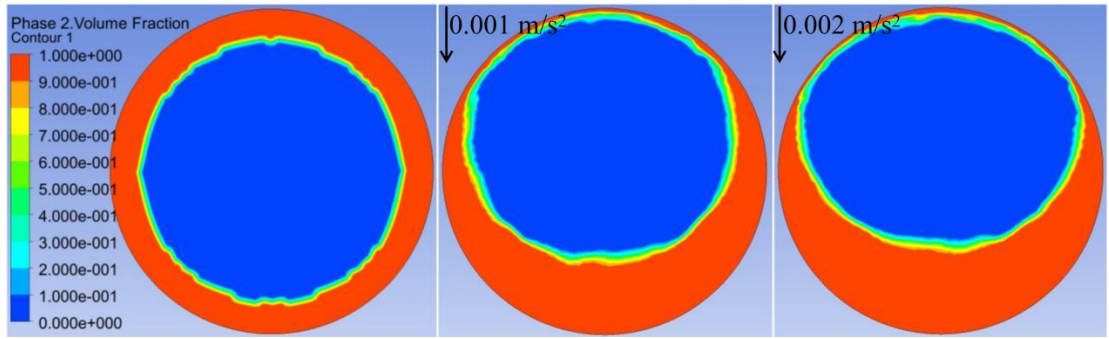

**Fig. 3** Liquid distribution in the half cross-section under different accelerations.

**2.3 Comparison between theoretical and numerical results**

Based on the expressions proposed above, two procedures are developed to predict the bubbles' profiles, which are named P1 and P2. P1 takes coordinates of Point *A* as inputs and outputs bubbles' profiles and volume. During the calculation process the boundary condition that the abscissas of the highest point on the profile and Point *B* is 0 is needed. In this situation the highest point on the predicted profile may not coincide with the original Point *O*. P2 takes bubble's volume as inputs and outputs bubbles' profiles. The shooting method is adopted in P2 and the highest point on the predicted profile coincides with Point *O*. During the calculation process, the range of the ordinate of Point *B* is estimated at first, and calculate bubbles' profiles and volume every 0.01 from the initial value of the ordinate range of Point *B*. When the difference between the calculated and given volume is less than 0.01%, the profile at the moment is the accurate prediction.

Figs. 4(a) and 4(b) shows comparison between predictions of these two procedures and numerical results. The small black dots in the upper part of the figures represent the points measured from numerical results. The red curves stand for the predicted profiles and the green curves stand for the circular profiles in absence of gravity. It can be seen that under a certain acceleration bubbles' profiles deviate from the circular shape. The black dots are in good agreement with the predicted red curves, which means numerical results are consistent with predictions of P1 and P2. Besides, in the predicted profile of P1, the ordinate of the highest point is smaller than 0, and the reason has been explained in last paragraph. Under different accelerations the static position of bubbles in containers is different. It's a pity that the method to predict the static position of bubbles hasn't been obtained in this study. Only bubbles' shapes can be predicted by using these two procedures, and this is also the reason that the original Point $O$ locates in the highest point of bubbles' profiles instead of the highest point of containers.

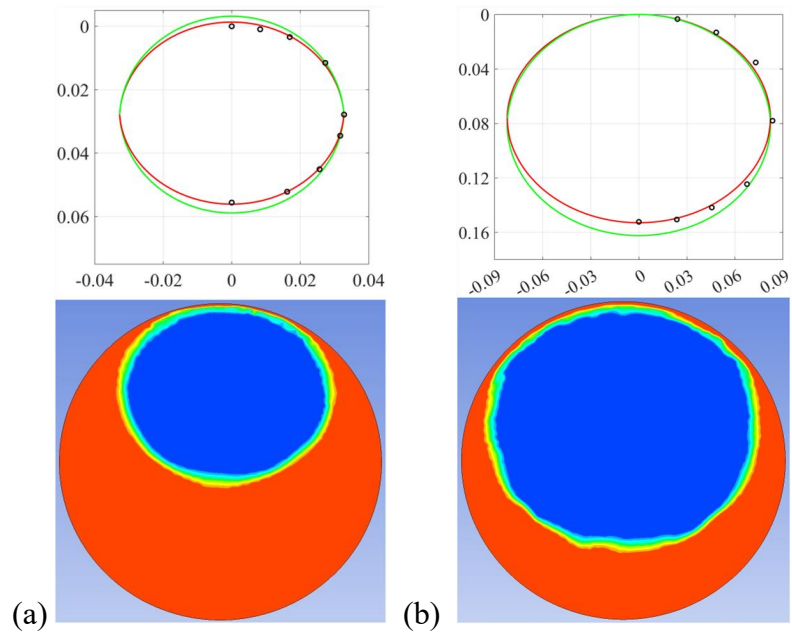

**Fig. 4** Comparison between theoretical predictions and numerical results. (a) Predictions of P1. $r_t$ = 50 mm, $g$ = 0.01m/s$^2$, (b) predictions of P2. $r_t$ = 100 mm, $g$ = 0.001m/s$^2$.

More numerical results are shown in Table 3. Different geometries, different bubble volume and different accelerations are all considered. Predictions of P1 and P2 are also presented. It can be seen that predictions of P1 and P2 are in good agreement with numerical results. Ratios of predicted to given volume and ratios of predicted to numerical $z_2$ are calculated. And values of the two ratios are mostly within ±105%.

For more intuitively comparison, these two ratios are presented in Fig. 5. The blue and black dots represent ratios of bubble volume and $z_2$ respectively. The deviation is mostly within ±5%, which verifies the accuracy of P1 and P2.

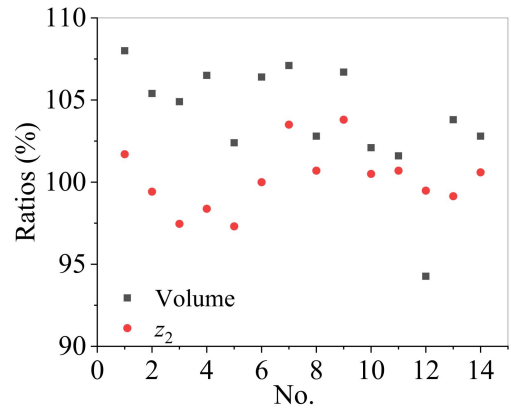

**Fig. 5** Ratios of predicted to given bubble volume and ratios of predicted to numerical $z_2$.

Table 3 Comparison between theoretical predictions and numerical results.

| No. | Model parameters | | | | Numerical results | | | | Predictions of P1 | | | Predictions of P2 | | | |
|---|---|---|---|---|---|---|---|---|---|---|---|---|---|---|---|
| | $r_t$ (mm) | $V_b$ (mm³) | $g$ (m/s²) | $Bo$ | $r_1$ (mm) | $z_1$ (mm) | $z_2$ (mm) | Ordinate of the highest point (mm) | $z_2$ (mm) | $V_b$ (mm³) | Ratios of predicted to given volume (%) | $r_1$ (mm) | $z_1$ (mm) | $z_2$ (mm) | Ratios of predicted to numerical $z_2$ (%) |
| 1. | 50 | 1.131e5 | 0.01 | 0.4187 | 32.80 | 27.86 | 55.99 | -0.8064 | 55.59 | 1.222e5 | 108.0 | 31.01 | 29.60 | 56.96 | 101.7 |
| 2. | 50 | 1.131e5 | 0.005 | 0.2093 | 31.92 | 28.89 | 58.22 | -0.1274 | 57.52 | 1.192e5 | 105.4 | 30.78 | 29.61 | 57.88 | 99.42 |
| 3. | 50 | 1.131e5 | 0.001 | 0.04187 | 31.49 | 29.53 | 60.55 | 0.052 | 58.94 | 1.186e5 | 104.9 | 30.48 | 29.64 | 59.01 | 97.46 |
| 4. | 50 | 2.681e5 | 0.005 | 0.3721 | 42.85 | 39.34 | 76.67 | 0.6911 | 76.88 | 2.854e5 | 106.5 | 41.70 | 38.38 | 75.43 | 98.38 |
| 5. | 50 | 2.681e5 | 0.002 | 0.1489 | 41.18 | 39.57 | 79.55 | 0.2947 | 78.54 | 2.746e5 | 102.4 | 41.02 | 39.08 | 77.41 | 97.31 |
| 6. | 100 | 2.681e5 | 0.002 | 0.1489 | 44.08 | 36.18 | 77.39 | -1.671 | 73.45 | 2.852e5 | 106.4 | 41.02 | 39.08 | 77.41 | 100.0 |
| 7. | 100 | 9.048e5 | 0.002 | 0.3350 | 65.54 | 55.04 | 110.2 | -2.072 | 110.7 | 9.693e5 | 107.1 | 61.96 | 57.36 | 114.0 | 103.5 |
| 8. | 100 | 9.048e5 | 0.001 | 0.1675 | 64.00 | 56.95 | 115.5 | -0.1594 | 113.3 | 9.301e5 | 102.8 | 61.31 | 58.42 | 116.3 | 100.7 |
| 9. | 100 | 2.145e6 | 0.002 | 0.5954 | 88.18 | 71.98 | 144.0 | -3.262 | 143.9 | 2.288e6 | 106.7 | 82.87 | 73.00 | 149.5 | 103.8 |
| 10. | 100 | 2.145e6 | 0.001 | 0.2977 | 83.53 | 77.97 | 152.2 | 0.864 | 153.5 | 2.190e6 | 102.1 | 82.25 | 76.11 | 152.9 | 100.5 |
| 11. | 100 | 3.054e6 | 0.001 | 0.3768 | 92.86 | 91.42 | 169.8 | 3.581 | 176.4 | 3.104e6 | 101.6 | 92.66 | 84.78 | 171.0 | 100.7 |

| 12. | 100 | 3.054e6 | 0.0006 | 0.2261 | 88.90 | 87.25 | 174.5 | 0.7991 | 173.8 | 2.879e6 | 94.27 | 92.05 | 86.59 | 173.6 | 99.48 |
| 13. | 200 | 1.716e7 | 0.0003 | 0.3573 | 169.9 | 151.1 | 301.1 | -2.337 | 300.3 | 1.781e7 | 103.8 | 166.8 | 150.0 | 298.5 | 99.14 |
| 14. | 200 | 2.443e7 | 0.0003 | 0.4521 | 188.8 | 177.4 | 330.2 | 2.911 | 344.1 | 2.511e7 | 102.8 | 188.1 | 166.7 | 332.3 | 100.6 |

Note: $Bo = \dfrac{\rho g r_g^2}{\sigma}$, where $r_g$ is the radius of the initial spherical bubble.

More predicted profiles of P2 are presented in Fig. 6. The dotted and solid curves represent the profiles of bubbles with $V_b$ = 2.145e6 mm$^3$ and $V_b$ = 3.054e6 mm$^3$ respectively. The black curves stand for profiles of bubbles in absence of gravity. When $Bo$ < 0.04, the profiles of bubbles are very close to those in absence of gravity and in this situation the effects of accelerations can be ignored. When $Bo$ > 0.2, the profiles are obviously different from those without gravity, and the bubbles become flatter. By using these two procedures bubbles' shape and volume with small $Bo$ numbers can be predicted accurately and quickly, which will be great helpful for liquid management and propellant residue gauging in space.

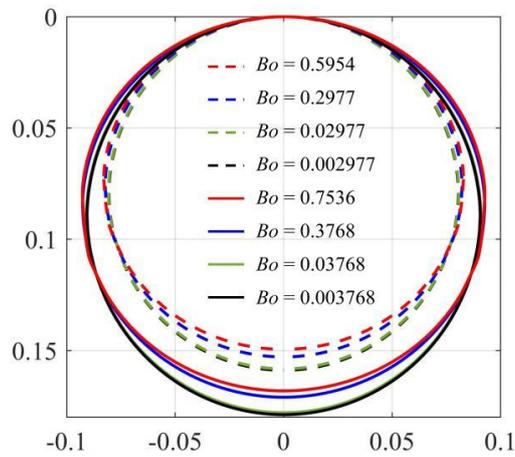

**Fig. 6** Predicted profiles of bubbles by using the second procedure.

Owing to the complex liquid distribution in space, it's hard to measure liquid residue in tanks directly. However, liquid residue can be inferred by measuring the bubble volume. The accurate measurement of several points on the bubble's profiles can be achieved with specific detection methods, especially the height and width of the bubble, that is, the coordinates of Points *A* and *B*. By using the first procedure, the bubble volume can be calculated with the coordinates of Point *A* as the input. In this process the influences of the acceleration on predictions of bubble volume can't be ignored. Fig. 7 shows ratios of predicted to given bubble volume of P1 vs $Bo$ numbers. P1 uses the abscissas of 50 mm, 100 mm, 200 mm, and 300 mm as inputs to predict the bubble volume under different accelerations (the ordinate of Point *A* only changes the bubble position, but has no effects on the bubble shape and volume). It can be seen that the smaller the bubble size, the more significant the influences will be. If the residual acceleration on satellites is ignored, there will be an unacceptable error on the measurement of liquid residue. For example, when $r_g$ = 100 mm and the $Bo$ number is 0.4652, ignoring the residual acceleration will cause a bubble volume measurement error of 5.90%.

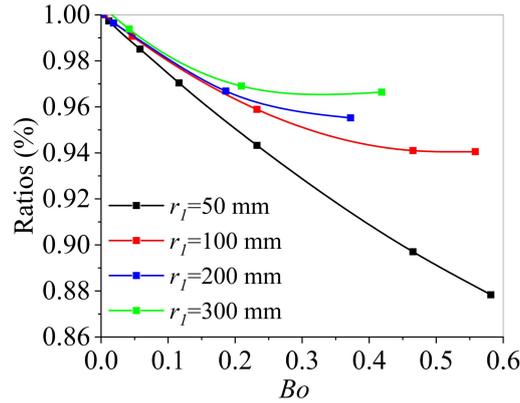

**Fig. 7** Ratios of predicted to given bubble volume vs $Bo$.

## 3. Liquid sinks to the bottom

### 3.1 Theoretical derivation

When the acceleration is large enough or the liquid volume is less enough, the liquid will all sink in the bottom of containers and form a curved liquid-gas interface as shown in Figs. 8(a)&8(b). When the $Bo$ number is small, the whole profile is a obviously curved, as shown in Fig. 8(a). But when the $Bo$>>1, the profile of the liquid-gas interface bends obviously only near the wall, and is approximately horizontal in the area far from the wall, as shown in Fig. 8(b).

The cylindrical coordinate is adopted for analysis. The $z$-axis coincides with the symmetry axis of the container, the $r$-axis is horizontal to the right, and the origin Point $O$ is at the lowest point of the free surface. The acceleration is along the negative direction of $z$-axis. The intersection of the profile and the wall is Point $C(r_3, z_3)$. The intersection of the profile and the $z$-axis is Point $D(0, 0)$, which is also the lowest point on the profile. The distance between Point $D$ and the lowest point of the container is $z_5$.

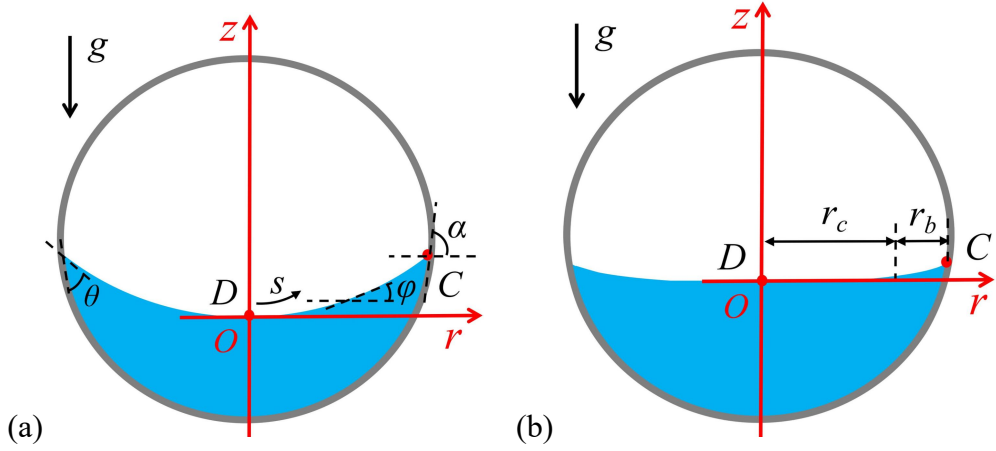

**Fig. 8** Liquid distribution under different *Bo* numbers. (a) Small *Bo* numbers, (b) large *Bo* numbers.

When *Bo* <<1, expressions of Curve *CD* can be derived by using the same perturbation method in Section 2, which are written as follows

$$\begin{aligned} Z &= Z_0 + \beta Z_1 \\ \varphi &= \varphi_0 + \beta \varphi_1 \\ A &= A_0 + \beta A_1 \\ B &= B_0 + \beta B_1 \end{aligned} \quad (3.1)$$

Where $\beta = \dfrac{\rho g r_t^2}{\sigma}$ is the *Bo* number.

$$\begin{aligned} Z_0 &= \frac{1}{\sin(\theta+\alpha)} - \frac{\sqrt{1-\sin^2(\alpha-\theta)r^2}}{\sin(\alpha-\theta)} \\ \varphi_0 &= \arcsin[r\sin(\alpha-\theta)] \\ A_0 &= 2\sin(\alpha-\theta) \\ B_0 &= 0 \end{aligned} \quad (3.2)$$

$$\begin{aligned} Z_1 &= \int_0^r \frac{1}{r\cos^3\varphi_0}[V_0 - V(r)]dr \\ \varphi_1 &= \frac{1}{r\cos\varphi_0}[V_0 - V(r)] \\ A_1 &= 2V_0 \\ B_1 &= 0 \end{aligned} \quad (3.3)$$

where

$$V_0 = \frac{\alpha-\theta}{2\sin^2(\alpha-\theta)} - \cot(\alpha-\theta) \quad (3.4)$$

When $Bo \gg 1$, to derive the expressions, the profile of liquid-gas interface is divided into two regions: one is the central region in which $\varphi$ is small, and the other is the boundary region close to the wall in which $\varphi$ increases rapidly to the boundary condition at Point C (subscript b represents the boundary region). Derive theoretical expressions of these two regions of profiles respectively, and make the expressions equal in the connecting point of these two regions, then the complete and uninterrupted expressions of profiles are obtained.

For convenience, the length quantities $r, z$ are nondimensionalize with the radius of container, $r_t$. By transforming Eq. 2.3, it can be obtained that

$$\frac{1}{R}\frac{d}{dR}\left\{\frac{RdZ/dR}{\left[1+(dZ/dR)^2\right]^{0.5}}\right\} - \beta Z - \gamma = 0 \tag{3.5}$$

where $R=r/r_t$, $Z=z/r_t$ and $\gamma = \dfrac{p_o r_t}{\sigma}$.

Eq. (3.5) becomes

$$\frac{1}{R_C}\frac{d}{dR_c}\left\{R_c \frac{dZ_c}{dR_c}\left[1+O(\varphi^2)\right]\right\} - \beta Z_c - \gamma = 0 \tag{3.6}$$

where the subscript c represents the central region. Combined with following boundary conditions,

$$R=0, Z=Z_4/r_t, \varphi=0 \tag{3.7}$$

The solution of Eq. (3.7) can be obtained

$$Z_c = \gamma\beta^{-1}\left[I_0\left(\beta^{\frac{1}{2}}R_c\right)-1\right]\left[1+O(\varphi^2)\right] \tag{3.8}$$

from which the relationship between $R_c$ and $\varphi$ can be obtained that

$$\varphi = \gamma\beta^{-0.5}I_1\left(\beta^{0.5}R_c\right)\left[1+O(\varphi^2)\right] \tag{3.9}$$

in which $I_0$ and $I_1$ are modified Bessel functions of order zero and one, respectively

Let $X$ be the variable in the boundary region, which is written as

$$X = \frac{1-R}{\beta^{-m}} \tag{3.10}$$

Substitute Eq. (3.10) into Eq. (3.5) leads to

$$\beta^m \frac{d\varphi}{dX}\cos\varphi + \frac{\sin\varphi}{1-\beta^{-m}X} - \beta Z - \gamma = 0 \qquad (3.11)$$

So *m* is determined to be 0.5.

Because $Bo \gg 1$, $\beta^{-0.5}$ is much smaller than 1. In the boundary region the perturbation method can be adopted by taking $\varepsilon = \beta^{-0.5}$ as the expansion parameter. Solutions of $Z(\varphi)$ and $R(\varphi)$ in the boundary region can be expressed as

$$Z_b(\varphi) = \varepsilon Z_1(\varphi) + \varepsilon^2 Z_2(\varphi) + O(\varepsilon^3) \qquad (3.12)$$

$$R_b(\varphi) = 1 - \varepsilon X_1(\varphi) - \varepsilon^2 X_2(\varphi) + O(\varepsilon^3) \qquad (3.13)$$

Combined with Eq. (3.10), the boundary condition at Point *C* is

$$\varphi = \alpha - \theta, \quad X_1(\varphi) = X_2(\varphi) = 0 \qquad (3.14)$$

In the central region $\varphi \ll 1$ and *Z* can be considered to be 0, so at the other end of the boundary region, the boundary condition is

$$\varphi = 0, \quad Z_1(\varphi) = Z_2(\varphi) = 0 \qquad (3.15)$$

Similar to Eqs. (2.11)&(2.12) in Section 2, Eq. (3.5) is also transformed into the parametric form, which is written as

$$\frac{dZ}{d\varphi} = \frac{\sin\varphi}{-\frac{\sin\varphi}{R} + \beta Z + \gamma} \qquad (3.16)$$

$$\frac{dR}{d\varphi} = \frac{\cos\varphi}{-\frac{\sin\varphi}{R} + \beta Z + \gamma} \qquad (3.17)$$

Substituting the first-order solutions of Eqs. (3.12)&(3.13) into Eqs. (3.16)&(3.17) and simplifying the results by ignoring the smaller terms yields the first-order equations

$$dZ_1/d\varphi = \sin\varphi/Z_1 \qquad (3.18)$$

$$dX_1/d\varphi = -\cos\varphi/Z_1 \qquad (3.19)$$

Combined with the boundary conditions, Eqs. (3.14)&(3.15), the solutions can be obtained as follows

$$Z_1 = 2\sin\frac{\varphi}{2} \qquad (3.20)$$

$$X_1 = \log\frac{\tan\frac{\varphi_w}{4}}{\tan\frac{\varphi}{4}} + 2\cos\frac{\varphi_w}{2} - 2\cos\frac{\varphi}{2} \qquad (3.21)$$

where $\varphi_w = \alpha - \theta$.

Substituting the second-order solutions of Eqs. (3.12)&(3.13) into Eqs. (3.16)&(3.17) and simplifying the results by ignoring the smaller terms yields the second-order equations

$$\frac{dZ_2}{d\varphi} = -\frac{\sin\varphi}{Z_1^2}Z_2 + \frac{\sin^2\varphi}{Z_1^2} \qquad (3.22)$$

$$\frac{dX_2}{d\varphi} = \frac{\cos\varphi}{Z_1^2}Z_2 - \frac{\sin\varphi\cos\varphi}{Z_1^2} \qquad (3.23)$$

Combined with Eqs. (3.14)&(3.15) and Eqs. (3.20)&(3.21), the solutions can be obtained as follows

$$Z_2 = \frac{2}{3}\frac{1-\cos^2\frac{\varphi}{2}}{\sin\frac{\varphi}{2}} \qquad (3.24)$$

$$X_2 = -\frac{2}{3}\sin^2\frac{\varphi_w}{2} - \frac{1}{6}\left(1+\cos\frac{\varphi_w}{2}\right)^{-1} + \frac{1}{2}\log\frac{\tan\frac{\varphi_w}{4}}{\tan\frac{\varphi}{4}} + \frac{2}{3}\sin^2\frac{\varphi}{2} + \frac{1}{6}\left(1+\cos\frac{\varphi}{2}\right)^{-1} \qquad (3.25)$$

Then the second-order perturbationo solutions of profiles in the boundary region are obtained.

By using the expansion parameter $\varepsilon$, the value of Eq. (3.8) at the connecting point is written as

$$Z_c = \varepsilon^2\gamma[I_0(R_c/\varepsilon)-1][1+O(\varepsilon^2)] \qquad (3.26)$$

The Bessel function in Eq. (3.26) may be approximated by its asymptotic expansion for large argument

$$Z_c = \varepsilon^2\gamma\frac{\exp(R_c/\varepsilon)}{(2\pi R_c/\varepsilon)^{0.5}}\left[1+\frac{\varepsilon}{8R_c}+O(\varepsilon^2)\right] \qquad (3.27)$$

The solution for the boundary region at the connecting point can be obtained

$$Z_1 = \varphi + O(\varphi^3) \tag{3.28}$$

$$X_1 = -\log\frac{\varphi}{4} + \log\tan\frac{\varphi_w}{4} + 2\cos\frac{\varphi_w}{2} - 2 + O(\varphi^2) \tag{3.29}$$

$$Z_2 = \frac{\varphi}{2} + O(\varphi^3) \tag{3.30}$$

$$X_2 = -\frac{1}{2}\log\frac{\varphi}{4} - \frac{2}{3}\sin^2\frac{\varphi_w}{2} - \frac{1}{6}\left(1+\cos\frac{\varphi_w}{2}\right)^{-1} + \frac{1}{2}\log\tan\frac{\varphi_w}{4} + \frac{1}{12} + O(\varphi^2) \tag{3.31}$$

Substituting Eqs. (3.28)-(3.31) yields

$$Z_b(\varphi) = \varepsilon\varphi + \frac{\varepsilon^2}{2}\varphi + O(\varepsilon^4) \tag{3.32}$$

$$1 - R_b(\varphi) = \varepsilon\left(1+\frac{\varepsilon}{2}\right)\left(-\log\frac{\varphi}{4} + \log\tan\frac{\varphi_w}{4}\right) - 2\varepsilon\left(1-\cos\frac{\varphi_w}{2}\right)$$
$$+ \varepsilon^2\left[\frac{1}{12} - \frac{2}{3}\sin^2\frac{\varphi_w}{2} - \frac{1}{6}\left(1+\cos\frac{\varphi_w}{2}\right)^{-1}\right] + O(\varepsilon^2) \tag{3.33}$$

Divide Eq. (3.33) by $\varepsilon$ and after exponentiation, Eq. (3.33) can be transformed to

$$\varphi = 4\exp\left\{\begin{array}{l}\log\tan\frac{\varphi_w}{4} - [1-R_b(\varphi)]/\varepsilon - 2\left(1-\cos\frac{\varphi_w}{2}\right) + \frac{1}{2}[1-R_b(\varphi)] \\ + \varepsilon\left[\frac{13}{12} - \cos\frac{\varphi_w}{2} - \frac{2}{3}\sin^2\frac{\varphi_w}{2} - \frac{1}{6}\left(1+\cos\frac{\varphi_w}{2}\right)^{-1} + O(\varepsilon)\right]\end{array}\right\} \tag{3.34}$$

Substituting Eq. (3.31) into Eq. (3.29) and after simplification leads to

$$Z_b(\varphi) = 4\varepsilon\left\{1+\frac{1}{2}[1-R_b(\varphi)] + \varepsilon\left[\frac{19}{12} - \cos\frac{\varphi_w}{2} - \frac{2}{3}\sin^2\frac{\varphi_w}{2} - \frac{1}{6}\left(1+\cos\frac{\varphi_w}{2}\right)^{-1}\right] + O(\varepsilon)\right\}$$
$$* \exp\left[-(1-R_b(\varphi))/\varepsilon + \log\tan\frac{\varphi_w}{4} - 2\left(1-\cos\frac{\varphi_w}{2}\right)\right]$$

$$\tag{3.35}$$

The result, after the relation $1-R_b(\varphi) = O(\varepsilon\log\varepsilon)$ has been used to expand (3.27) is transformed to

$$Z_c = (2\pi)^{-\frac{1}{2}}\gamma\varepsilon^{\frac{5}{2}}\exp(R_c/\varepsilon)\left[1+\frac{1}{2}(1-R_c) + \frac{1}{8}\varepsilon + o(\varepsilon)\right] \tag{3.36}$$

From which the expressions of $\gamma$ can be derived

$$\gamma = e^{-1/\varepsilon}\varepsilon^{-3/2}\gamma_0\left[1+\varepsilon\gamma_1+o(\varepsilon)\right] \tag{3.37}$$

where

$$\gamma_0 = 4(2\pi)^{\frac{1}{2}}\tan\frac{\varphi_w}{4}\exp\left[-2\left(1-\cos\frac{\varphi_w}{2}\right)\right] \tag{3.38}$$

$$\gamma_1 = \frac{35}{24}-\cos\frac{\varphi_w}{2}-\frac{2}{3}\sin^2\frac{\varphi_w}{2}-\frac{1}{6}\left[1+\cos\frac{\varphi_w}{2}\right]^{-1} \tag{3.39}$$

Eqs. (3.37)-(3.39) shows that $\gamma$ is smaller than terms of order one, which verifies that it can be ignored compared with ohter terms in Eq. (3.5) when $Bo \gg 1$.

**3.2 Numerical simulation**

SF 2 and SF 5 are adopted in the simulation and their properties are listed in Table 4. Numerical settings are the same as those in Table x. Different geometries, different liquid contact angles, different accelerations and different liquid volume are considered. Fig. 9 shows initial and static liquid distribution under different accelerations in the half cross-section. The acceleration and direction are also presented in the figure. In the beginning liquid is all in the bottom and the liquid-gas interface is flat. When $g = 0.01$ m/s² the static liquid-gas interface curved obviously. The $Bo$ number is 0.4293 and the surface tension plays a major role. When $g = 1$ m/s² the static liquid-gas interface is only curved in the small region near the wall. The $Bo$ number is 42.93 and the acceleration plays a greater role. The coordinates of Points $C$ and $D$ are measured from the numerical results and presented in Table 5.

Table 4 Fluid properties (25°C)

| Fluid | $\mu$ (kg/(m*s)) | $\rho$ (kg/m³) | $\sigma$ (N/m) | $v$ ($10^{-6}$ m²/s) |
|---|---|---|---|---|
| Air | 1.789e-5 | 1.225 | / | 1.460e-5 |
| SF 2 | 0.001746 | 873 | 0.0183 | 2 |
| SF 5 | 0.004575 | 915 | 0.0197 | 5 |

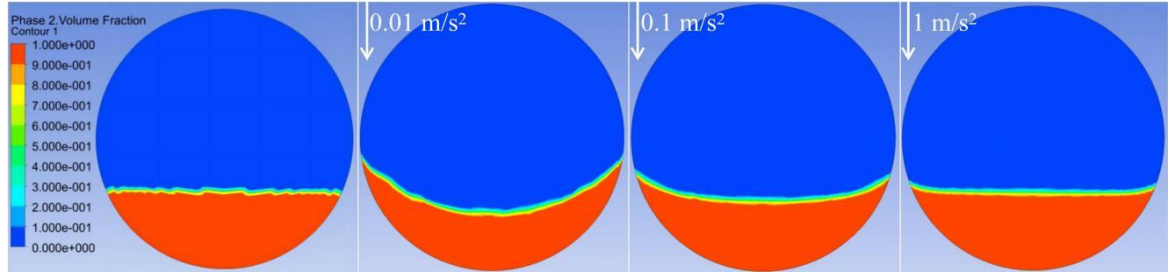

**Fig. 9** Liquid distribution in the half cross-section. $r_t$ =50 mm, $\theta = 30°$, $V_l$ = 1.131e5 mm$^3$.

### 3.3 Comparison between theoretical and numerical results

Based on the theoretical expressions proposed in Section 3.1, four procedures are developed to predict profiles of the free surfaces, which are named P3~P6 respectively. P3 and P5 take the coordinates of Point *C* as the input and predict the free surface and liquid volume under small and large *Bo* numbers respectively. P4 and P6 take the liquid volume as the input and predict the free surface under small and large *Bo* numbers respectively. Coordinates of several points on profiles in numerical results are measured and plotted with predictions of P3 and P5, as shown in Fig. 10. The black circles represent numerical results. The red and blue curves stand for the profiles under small and large *Bo* numbers. The circles are in good agreement with predictions of P3 and P5, which verifies the accuracy of the procedures. More comparison between numerical results and predictions of these procedures is presented in Table 5. Ratios of predicted to numerical results are mostly within 1±5%, which verifies the accuracy of our procedures.

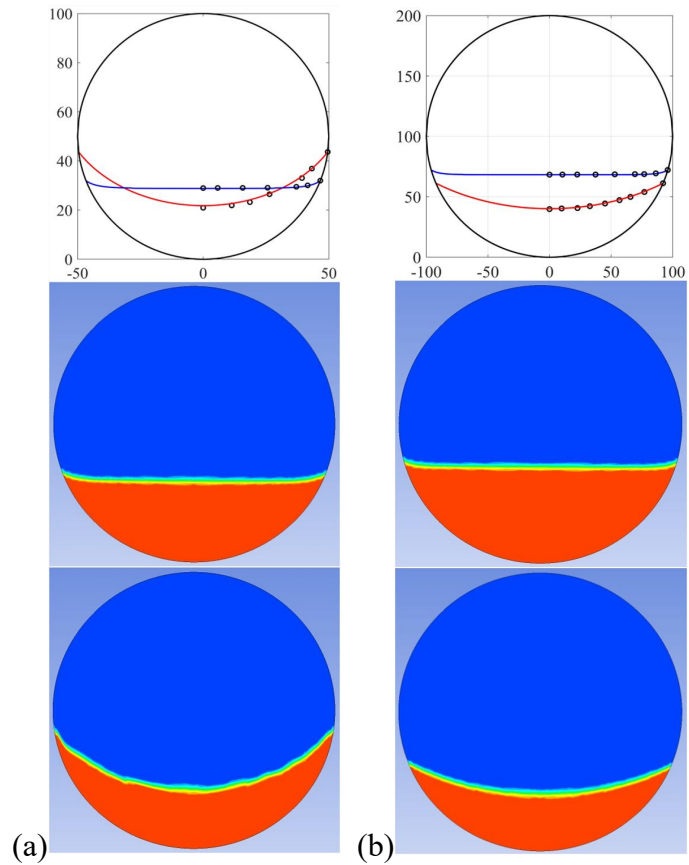

(a) (b)

**Fig. 10** Comparison between theoretical predictions and numerical results. (a) SF2, $r_t$ =50 mm, $\theta = 30°$, $V_l = 1.131e5$ mm$^3$, (b) SF5, $r_t$ =100 mm, $\theta = 40°$, $V_l = 6.545e5$ mm$^3$ and 1.180e6 mm$^3$ respectively.

More predictions are shown in Fig. 11(a)~11(d). Figs. 11(a)&11(b) shows predictions of P3 and P5, and the inputs of $z_3$ is 50 mm and 100 mm respectively. When $g$ is smaller than 0.01 m/s$^2$, $Bo$ <<1 and the profiles curved obviously. When $g$ is bigger than 0.1 m/s$^2$, $Bo$ >>1, and the profiles are almost flat in most regions and only curved obviously in the region near the wall, which is consistent with the theoretical analysis. It can be seen that under different accelerations the liquid volume differs a lot with the same meniscus height on the wall. A new method of liquid residue measurement method is to identify the meniscus position on the wall by its thermal response characteristics so as to evaluate the liquid residue. The residual acceleration in space and small acceleration caused by satellites maneuver will have great effects on the profiles. By using these procedures the effects of accelerations can be predicted and the measurement accuracy of liquid residue can be ensured. Figs.

11(c) and 11(d) shows predictions of P4 and P6, and the inputs of $V_l$ is 6.545e5 mm$^3$ and 2.094e6 mm$^3$ respectively. Likewise, the meniscus height on the wall when $Bo$<< 1 is much higher than that when $Bo$ >> 1. When $Bo$<< 1, the profiles under different accelerations differs a lot. But when $Bo$ >> 1, the profiles only curved obviously in the small region near the wall regardless of accelerations. Because some small quantities are ignored in the derivation process, P3~P6 are not applicable when the $Bo$ number is of order one.

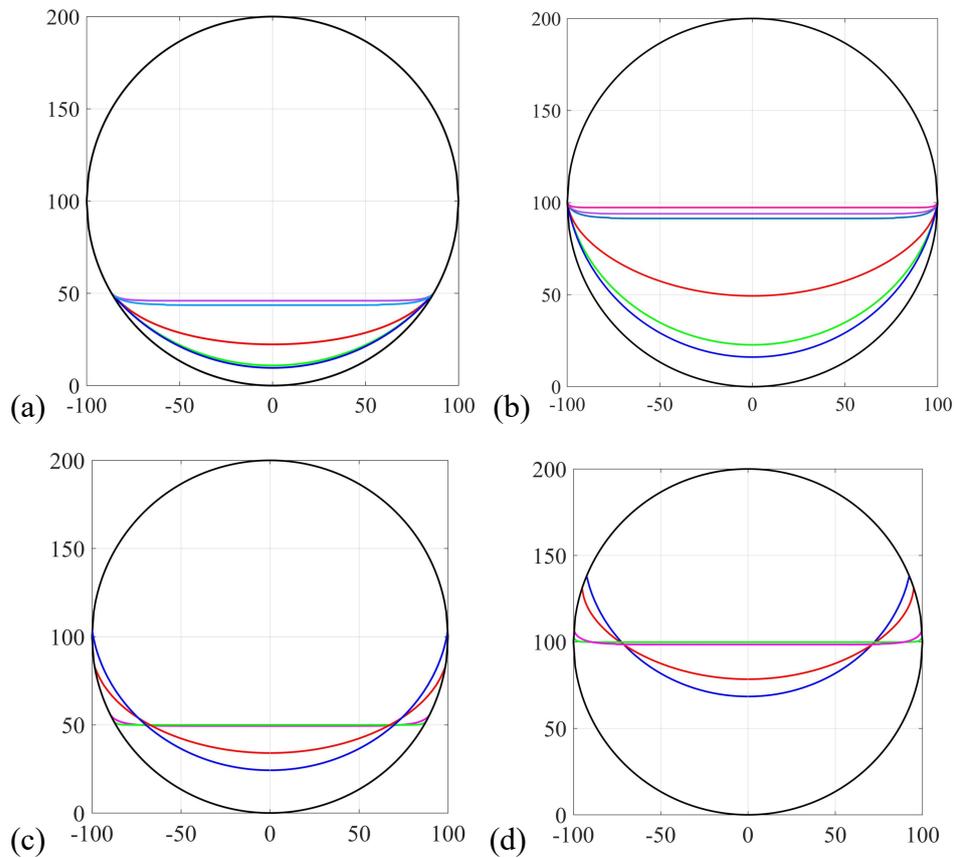

Fig. 11 Theoretical predictions of the profiles. The liquid is SF5 and the contact angle is 10°. (a) Predictions of P3 and P5. $z_3$ = 50 mm, $g$ = 0, 0.001, 0.01, 0.4 and 1 m/s$^2$ respectively. (b) Predictions of P3 and P5. $z_3$ = 100 mm, $g$ = 0, 0.001, 0.005, 0.5, 1 and 5 m/s$^2$ respectively. (c) Predictions of P4 and P6. $V_l$ = 6.545e5 mm$^3$. $g$ = 0.001, 0.005, 0.5 and 5 m/s$^2$ respectively. (d) Predictions of P4 and P6. $V_l$ = 2.094e6 mm$^3$. $g$ = 0.001, 0.005, 0.5 and 5 m/s$^2$ respectively.

Recently a propellant residue measurement technique, which evaluates propellant residue by identifying the meniscus position on the wall through the thermal response of the wall, has been developed. It's meaningful to reveal the effects of accelerations and liquid contact anlge on the evaluation of liquid volume. Liquid volume calculated by P3 and P5 with the same inputs of $z_3$ under different accelerations are shown in Fig. 12. When $\alpha - \theta > 0°$, liquid volume increases with the increase of the acceleration. And the most significant change of liquid volume is when $g<1$ m/s$^2$. Because the profiles changes a lot when $g<1$ m/s$^2$ as shown in Fig. 11(a) ~11(d). But when $\alpha - \theta <0°$, liquid volume decreases with the increase of the acceleration. This is because when $\alpha - \theta <0°$, the free surfaces are convex and the rules of liquid volume change is contrary to those when $\alpha - \theta >0°$.

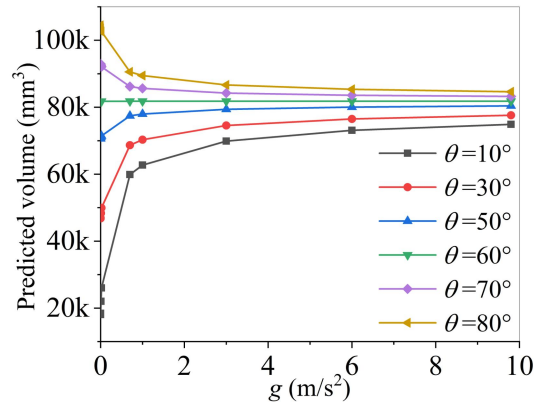

**Fig. 12** Liquid volume calculated by P3 and P5 with the same inputs of $z_3$ under different accelerations. Liquid is SF2. $r_t$ =50 mm, $z_3$ = 25 mm, $\theta$ = 10°, 30°, 50°, 60°, 70°, 80°.

Table 5 Comparison between theoretical predictions and numerical results.

| No. | Liquid | Contact angle $\theta$ (°) | Model parameters | | | | | Numerical results | | | Predictions of P3 and P5 | | | Predictions of P4 and P6 | | |
|---|---|---|---|---|---|---|---|---|---|---|---|---|---|---|---|---|
| | | | $r_t$ (mm) | $V_l$ (mm³) | $g$ (m/s²) | $Bo$ | $r_3$ (mm) | $z_3$ (mm) | $z_5$ (mm) | $z_5$ (mm) | $V_l$ (mm³) | Ratios of predicted to given volume (%) | $z_3$ (mm) | $z_5$ (mm) | Ratios of predicted to given $z_3$ (%) |
| 1. | SF2 | 30 | 50 | 1.131e5 | 1 | 42.93 | 46.60 | 31.97 | 28.96 | 28.81 | 1.088e5 | 96.20 | 32.68 | 29.46 | 102.2 |
| 2. | SF2 | 30 | 50 | 1.131e5 | 0.1 | 4.293 | 48.03 | 36.23 | 25.97 | 25.90 | 1.107e5 | 97.88 | 36.69 | 26.23 | 101.3 |
| 3. | SF2 | 30 | 50 | 1.131e5 | 0.01 | 0.4293 | 49.56 | 43.66 | 20.84 | 21.81 | 1.183e5 | 104.6 | 42.43 | 21.28 | 97.18 |
| 4. | SF2 | 30 | 50 | 1.131e5 | 0.001 | 0.04293 | 49.65 | 44.48 | 19.25 | 19.63 | 1.135e5 | 100.4 | 42.39 | 19.06 | 95.30 |
| 5. | SF2 | 60 | 50 | 1.131e5 | 0.01 | 0.4293 | 46.37 | 31.41 | 27.70 | 28.31 | 1.117e5 | 98.76 | 31.81 | 28.51 | 101.3 |
| 6. | SF2 | 60 | 50 | 1.131e5 | 0.001 | 0.04293 | 46.36 | 31.40 | 27.56 | 28.09 | 1.112e5 | 98.32 | 31.78 | 28.28 | 101.2 |
| 7. | SF2 | 60 | 50 | 1.131e5 | 0.0001 | 0.004293 | 46.38 | 31.40 | 27.58 | 28.09 | 1.111e5 | 98.23 | 31.81 | 28.28 | 101.2 |
| 8. | SF2 | 20 | 100 | 4.356e5 | 0.002 | 0.2110 | 94.67 | 67.94 | 27.06 | 26.05 | 4.734e5 | 108.7 | 68.41 | 26.11 | 107.0 |
| 9. | SF5 | 40 | 100 | 6.545e5 | 0.1 | 13.47 | 88.14 | 52.88 | 47.45 | 46.96 | 6.259e5 | 95.63 | 54.29 | 48.13 | 102.7 |
| 10. | SF5 | 40 | 100 | 6.545e5 | 0.002 | 0.2694 | 92.14 | 61.23 | 39.85 | 40.16 | 6.451e5 | 98.56 | 62.62 | 40.74 | 102.3 |
| 11. | SF5 | 40 | 100 | 6.545e5 | 0.0002 | 0.02694 | 92.13 | 61.28 | 39.04 | 39.06 | 6.347e5 | 96.97 | 62.62 | 39.58 | 102.2 |

| 12. | SF5 | 40 | 100 | 1.180e6 | 0.5 | 99.80 | 95.96 | 72.22 | 68.34 | 68.30 | 1.145e6 | 97.03 | 73.51 | 69.50 | 101.8 |
| 13. | SF5 | 40 | 100 | 1.180e6 | 0.1 | 19.96 | 97.06 | 76.14 | 66.92 | 66.51 | 1.155e6 | 97.88 | 77.12 | 67.34 | 101.3 |
| 14. | SF5 | 30 | 300 | 1.767e7 | 0.5 | 606.5 | 260.5 | 151.4 | 147.8 | 144.4 | 1.653e7 | 93.55 | 153.4 | 149.9 | 101.3 |
| 15. | SF5 | 30 | 300 | 1.767e7 | 0.1 | 121.3 | 261.9 | 154.0 | 145.4 | 146.0 | 1.699e7 | 96.15 | 157.4 | 149.2 | 102.2 |
| 16. | SF5 | 30 | 300 | 1.767e7 | 0.05 | 60.65 | 263.9 | 157.5 | 144.4 | 145.8 | 1.713e7 | 96.94 | 160.3 | 148.4 | 101.8 |

Note: $Bo = \dfrac{\rho g r_e^2}{\sigma}, r_e = \left(\dfrac{3V_l}{4\pi}\right)^{1/3}$.

# 4. Conclusions

Profiles of bubbles suspended in the liquid and liquid-gas interfaces when liquid all sinks in the bottom under different accelerations are explored deeply in this study, and theoretical expressions of the profiles are derived by using the perturbation method. Six procedures are developed based on these expressions, which are named P1~P6. They can be divided into two types. One types of procedures takes coordinates of endpoints of the profiles as inputs, and the other types of procedures takes liquid volume or gas volume as inputs. Numerical simulation by considering different volume, different accelerations, different contact angles and different geometries are performed with the VOF method and numerical results are in good agreement with predictions of P1~P6.

Second-order perturbation solutions of profiles of bubbles with small *Bo* numbers are derived by balancing the capillary pressure with the hydrostatic pressure caused by the accelerations. The bubble is spherical in the absence of gravity, and under the effects of accelerations it will become flatter. The bigger the acceleration, the more flatter it will be until all liquid sinks to the bottom. Liquid resudue can be evaluated by measuring several points on the bubble's profile and predicting the bubble volume with P1. Effects of accelerations on bubbles' shape must be considered, otherwise it will cause a volume measurement error of about 10%.

When liquid all sinks in the bottom, profiles of liquid-gas interfaces when *Bo* << 1 can be seen a part of profiles of bubbles. When *Bo* >> 1, profiles are divided into the central region and the boundary region. In the central region profiles are nearly straight and the dip angle $\varphi$ is much smaller than 1. In the boundary region profiles curved obviously. Most parts of the profile belong to the central region, and the width and height of the boundary region are both the same order of as $\beta^{-0.5}$.

Predictions of liquid volume with the same liquid meniscus height as inputs differs a lot under different accelerations. The most significant change of liquid volume is when *Bo* << 1. Besides, when *α* - *θ* > 0°, liquid volume increases with the increase of the acceleration. But when *α* - *θ* < 0°, liquid volume decreases with the increase of the acceleration. This is because when $α - θ <0°$, the free surfaces are convex and the rules of liquid volume change is contrary to those when $α - θ >0°$. When evaluating liquid residue through the thermal response of the wall, the effects of residual accelerations in space and accelerations caused by satellites maneuver must be considered.

Liquid management and residue measurement in space are of great significance for the long-term stable operation of spacecrafts. The theoretical models and calculation procedures proposed in this paper are of great significance, which can provide theoretical basis and design tools for improving the evaluation accuracy of liquid residue and fine management of fluid in space.

## Acknowledgements

The research is supported by the National Key R&D Program of China (2020YFC2201100).